\documentclass{article} 
\usepackage[twocolumn]{emulateapj}
\usepackage{psfig}

\newcommand{\bc}{\begin{center}}
\newcommand{\ec}{\end{center}}
\newcommand{\bl}{\begin{flushleft}}
\newcommand{\el}{\end{flushleft}}

\newcommand{\beq}{\begin{equation}}
\newcommand{\eeq}{\end{equation}}

\begin{document}

\submitted{Accepted for publication in ApJL}

\title{On the radio-to-X-ray light curves of SN 1998bw and GRB 980425}

\author{Koichi IWAMOTO\altaffilmark{1,2}}

\altaffiltext{1}{Department of Astronomy, University of Tokyo, Bunkyo
Ward, Tokyo 113-0033, Japan}

\altaffiltext{2}{Department of Physics, Nihon University, Chiyoda
Ward, Tokyo 101-8308, Japan}

\received{1998, October 26}

\begin{abstract} 

We calculate radio-to-X-ray light curves for afterglows caused by
non-thermal emission from a highly relativistic blast wave, which is
inferred from the gamma-ray flux detected in GRB 980425 and from the
very bright radio emission detected in SN 1998bw.  We find that the
observed gamma-ray and radio light curves are roughly reproduced by
the synchrotron emission from a relativistic fireball.  The optical
flux predicted for the non-thermal emission is well below that of the
thermal emission observed for SN 1998bw so that it will not be seen at
least for a few years.  The model predicts the X-ray flux just above
the detection limit of {it BeppoSAX} for the epoch when it was pointed
to the field of GRB980425. Therefore, the nondetection of X-ray and
optical afterglows is consistent with the model. The models presented
here are consistent with the physical association between SN 1998bw
and GRB980425, and lend further support to the idea that this object
might correspond to an event similar to the ``hypernova'' or
``collapsar'' -- events in which the collapse of a massive star forms
a rotating black hole surrounded by a disk of the remnant stellar
mantle.
\end{abstract}



\section{Introduction}

The discovery of the peculiar supernova SN 1998bw in the error box of
the \(\gamma\)-ray burst GRB 980425 raised the possibility that at
least some classes of \(\gamma\)-ray bursts may originate from
supernovae(Galama et al. 1998, Kulkarni et al. 1998).  The optical
properties of SN 1998bw suggest that it was an explosion of a massive
star with a kinetic energy 10 times larger than usual supernovae
(Iwamoto et al.  1998, Woosley, Eastman, \& Schmidt 1998). The radio
light curves of SN 1998bw indicate the existence of a relativistic
blast wave associated with the supernova (Kulkarni et al. 1998).  The
\(\gamma\)-ray burst GRB 980425 was an unremarkable event in terms of
spectral and temporal properties.  The time of occurrence of the burst
coincides with that of supernova to within (+0.7, -2.0) days, and the
supernova was found in the 8' error box of GRB 980425 determined by
{\it BeppoSAX}.  No further evidence has been established so far to
support the physical association of these two events, however, the
chance probability of finding a supernova in the error box is
estimated to be relatively small, \( \sim 9 \times 10^{-5}\) (Galama
et al. 1998).  Provided that GRB 980425 was really associated with SN
1998bw, the energy radiated in $\gamma$-rays turns out to be \(\sim 8
\times 10^{47}\) erg, assuming isotropic emission, which is 4 orders
of magnitude smaller than in other bursts whose distances have been
estimated so far(Galama et al. 1998).

The possible supernova connection of the \(\gamma\)-ray burst reminds us
of ``hypernova'' (Paczy\'nski 1998) or ``collapsar'' (Woosley et
al.1998; MacFadyen \& Woosley 1998) scenarios for \(\gamma\)-ray
bursters, in which the collapse of a very massive star is involved.
The scenario is that the collapse of the star results in the formation
of a system composed of a Kerr black hole and a disc of the remnant
stellar mantle.  Then, a required amount of energy may be extracted
from the system by either neutrino annihilation or
megneto-hydrodynamic effect, i.e., the Blandford-Znajek
mechanism(Blandford \& Znajek 1977), eventually producing a
relativistic jet and a bulk mass ejection.  In these scenarios, the
high energy photons from the jet may be observed as a \(\gamma\)-ray
burst and the bulk mass with a slower expansion speed would be seen as
a supernova-like object like SN 1998bw. Therefore, SN 1998bw and
GRB980425 provide a good opportunity to test this class of models for
gamma-ray bursters.  In this Letter, we combine the observational data
in different wavelength bands and try to see if they can be consistent
with the hypernova or collapsar scenarios.  We briefly give a
description of a fireball model that we apply in \S 2. Then we compare
the model prediction with the data in \S 3 and discuss possible
interpretations and their implications in \S 4. Concluding remarks
will be given in \S 5.

\section{Blast wave model}

We apply a simple fireball model for the blast wave dynamics, as in
M\'esz\'aros \& Rees(1997) and Wijers, Rees, \& M\'esz\'aros(1997).
The blast wave with an initially large Lorentz factor \(\Gamma\)
decelerates as it sweeps up external matter, giving rise to
non-thermal emissions at a decreasing characteristic frequency. The
synchrotron radiation from electrons accelerated near the shock front
is one of the most viable emission mechanisms, unless the density is
too low and the energy transfer from protons to electrons does not
occur sufficiently quickly.  The purpose of this Letter is not to give
a unique and comprehensive model but to check the consistency of the
physical association of GRB980425 and SN 1998bw. Therefore, we try a
simple fireball model similar to those applied to GRB970228 in Wijers
et al.(1997).

Recently, Rees \& M\'esz\'aros(1998) suggested the possibility that
the slower parts of ejecta would catch up with the decelerated blast
wave and may reenergize it.  The energy could be supplied to the blast
wave during its propagation, resulting in afterglows that could be
even more powerful than the $\gamma$-ray burst itself.  In our
analysis, we also assume a power-law \(\Gamma\) evolution so that
\(\Gamma \propto r^{-n}\) as in Rees \& M\'esz\'aros(1998), which is
realized in self-similar solutions(Blandford \& McKee 1976).  Since
the blast wave radius \(r\) is related to the observer time \(t\) by
\(r \sim 2 c t \Gamma^{2}\), the radius and the Lorentz factor are
given by

\begin{equation} 
r \propto t^{1/(2n+1)}, \hspace{1cm} \Gamma \propto t^{-n/(2n+1)}.
\end{equation}

The comoving frame synchrotron intensity is given by \( I'_{\nu}
\propto \nu^{\alpha}\) for \(\nu < \nu'_{\rm m}\) and \( I'_{\nu}
\propto \nu^{\beta}\) for \(\nu > \nu'_{\rm m}\), where \(\nu'_{\rm
m}\) is the break frequency \( \nu'_{\rm m} \propto B' \gamma_{\rm e,
m}^{2}\) and \( \alpha = 1/3 \), \(\beta = -(p-1)/2 \), and \(p\) is
the index of the energy spectrum of nonthermal electrons \(N_{\rm
e}(\gamma_{\rm e}) \propto \gamma_{\rm e}^{-p}\).  The typical Lorentz
factor of electrons is given by \(\gamma_{\rm e, m} \sim (m_{\rm
p}/m_{\rm e}) \cdot \Gamma \propto \Gamma\) so that \(\nu'_{\rm m}
\propto \Gamma^{3}\). The equipartition between magnetic field and
internal energies, \(B'^{2}/8 \pi \sim \Gamma^{2} n_{\rm ex} m_{\rm p}
c^{2}\), yields \(B' \propto \Gamma\). (Primes indicate values
evaluated in the comoving frame.)  The observed flux at the break
frequency is given by \( F_{\nu_{\rm m}} \propto (c t \Gamma)^{2}
\Gamma^{3} I'_{\nu'_{\rm m}} \propto t^{2} \Gamma^{5} I'_{\nu'_{\rm
m}}\).  The comoving intensity at the break frequency is given by \(
I'_{\nu_{\rm m}} \propto n'_{\rm e} B'^{2} \gamma_{\rm e, m}^{2}
\Delta r/(B'\gamma_{\rm e, m}^{2}) \propto n'_{\rm e} B' \Delta r
\propto \Gamma \cdot \Gamma \cdot r \Gamma^{-1} = \Gamma r \).

Therefore, the observed break frequency and the flux at the break
frequency evolve as

\begin{equation}
\nu_{\rm m} \sim \Gamma \nu'_{\rm m} \propto \Gamma^{4} \propto
t^{-4n/(2n+1)}
\end{equation}

\noindent
and

\begin{equation}
F_{\nu_{\rm m}} \propto t^{2} \Gamma^{6} r \propto t^{-(2n-3)/(2n+1)},
\end{equation}

\noindent
respectively.  

The index \(n\) is determined by equating the ram pressures on the
shocked region from the forward shock \(p_{\rm f}\) and from the
reverse shock \(p_{\rm rev}\) (Rees \& M\'esz\'aros 1998).  We assume,
for simplicity, that the mass contained in the layers outside the
shell with \(\Gamma = \Gamma_{\rm f}\), \(M(\Gamma > \Gamma_{\rm
f})\), is given by
 
\begin{equation}
M(\Gamma > \Gamma_{\rm f}) \propto \Gamma_{\rm f}^{-s},
\end{equation}

\noindent
and a power-law density structure of external medium

\begin{equation}
\rho_{\rm ext} \propto r^{-t}.
\end{equation}

The mass that has caught up with the contact discontinuity when the
blast wave reaches radius \(r\) is given by \(M_{\rm r} \propto
r^{ns}\).  Then we have \(p_{\rm f} \propto d M_{\rm r} / d r \propto
r^{ns-1}\) and \(p_{\rm rev} \propto \rho_{\rm ext} \Gamma^{2}
r^{2}\). In the latter, we assume the blast wave is adiabatic.  From
\(p_{\rm f} \sim p_{\rm rev}\), we find the index to be 

\begin{equation}
n =\frac{3-t}{s+2}.
\end{equation}

For a uniform distribution of external matter ( \(t = 0\) ) and a
shell with a unique \(\Gamma\) ( \( s= 0 \)), we have \(n = 3/2\),
which reproduces the result obtained in Wijers et al.(1997): \( r
\propto t^{1/4}, \Gamma \propto t^{-3/8}, \nu_{m} \propto t^{-3/2},
F_{\nu_{\rm m}} \propto t^{0} = \) const.

The monochromatic light curve is then given, if no beaming occurs, by

\[
F_{\nu} = F_{\nu_{\rm m}} \left(\frac{\nu}{\nu_{\rm
m}}\right)^{\alpha} \propto t^{\delta}, \hspace{0.5cm}
\delta = \frac{-2n+3+4n \alpha}{2n+1},
\hspace{0.5cm} t < t_{\nu} 
\]

\noindent
and

\begin{equation}
F_{\nu} = F_{\nu_{\rm m}} \left(\frac{\nu}{\nu_{\rm
m}}\right)^{\beta} \propto t^{\delta'}, \hspace{0.5cm}
\delta' = \frac{-2n+3+4n \beta}{2n+1},
\hspace{0.5cm} t > t_{\nu},
\end{equation}

\noindent
where \(t_{\nu}\) is the time of the light curve break at frequency
\(\nu\), which satisfies

\begin{equation} 
\frac{t_{\nu'}}{t_{\nu}} = \left(\frac{\nu}{\nu'}\right)^{1/2+1/4n}, 
\end{equation}

\noindent
as seen from equations (2) and (7).

\section{Comparison with the Data}

Table 1 lists the reports on the \(\gamma\)- and X-ray detections in
the Wide-Field Camera(WFC) error box of GRB980425~(Pian et al. 1998a,
1998b). We converted these data to the fluxes in mJy ($=10^{-26}$ erg
s$^{-1}$ cm$^{-2}$ Hz$^{-1}$) and plotted them in Figure 1 with open
and filled circles, respectively.  The observed optical light curve of
SN 1998bw in visual band(\(\lambda \sim 5500\) \AA) (Galama et
al. 1998) and the radio radio light curve at 6 cm ( Kulkarni et
al. 1998) are also plotted with open and filled squares, respectively.
A theoretical light curve of model CO138 (Iwamoto et al. 1998) for SN
1998bw, which is the thermal emission from a subrelativistic ejecta
heated by radioactivity, is shown with a solid line.  Figure 1 also
shows a set of afterglow light curves in X-rays($ h \nu = $ 5keV)({\it
dashed line}), optical(visual band, which is at approximately $\lambda
\sim$ 5,500 \AA; {\it dash-dotted line}), and radio ($\lambda $= 6 cm;
{\it dotted line}). It is seen that the radio light curve agrees with
the model prediction, although it shows a complicated behaviour before
and around the peak.

The gap of the peak fluxes between \(\gamma\)-ray and radio bands
requires that \(F_{\nu_{\rm m}}\) should increase as a function of
time. The slope of \(F_{\nu_{\rm m}}\) is chosen to be 1/3 as shown in
Figure 1, which requires \( n=1\), thus \( s+t = 1\).  In order for
the X-ray flux to fall below the detection limit at $t = 10^{4.93}$ s
= 1 d, the X-ray light curve should decay faster than \( t^{-1.2}\),
which means \( \beta < - 1.15\) if \( n = 1\) as seen from equation
(7). We choose \( \beta = - 1.15\) to draw the afterglow light curves
in Figure 1. Owing to the unknown evolution of random magnetic field
strength \(B'\) or due to the other emission processes involved such
as self-absorptions and the inverse Compton scattering, the spectrum
shape is uncertain and several breaks are even expected to appear in
the spectrum(Piran 1997). Therefore, we simply assume \(\alpha = 0\)
as adopted in Wijers et al.(1997) for our analysis in this Letter.

From equation (8) with \(n=1\), one expects the time of the breaks at
the X-ray, optical, and radio bands as \( t_{\rm X} = 10^{0.98}
t_{\gamma}, t_{\rm optical} = 10^{3.5} t_{\gamma}, t_{\rm radio} =
10^{7.2} t_{\gamma}\), respectively.  The flux of the optical
afterglow is expected to be well below the observed optical flux of SN
1998bw; thus, it is no wonder that it has not been detected.  The
model predicts that the optical afterglow will not be seen at least
for a few years, and then it will be below detection limits even with
a largest scale of telescopes. In contrast, the radio flux is still at
a level close to its peak, and monitoring the manner of its further
decay is crucial to distinguish the nature of the blast wave.  The
X-ray flux detected by {\it BeppoSAX} at the time of the burst is
significantly large compared with the \(\gamma\)-ray flux itself so
that it seems hard to explain.

From the radio properties of SN 1998bw, Kulkarni et al. (1998)
inferred that there exists a relativistic shock with \(\Gamma \sim
1-2\) at \( t = 10^{6.5}\) s that is responsible for the radio
emission.  Equation (1) predicts \(\Gamma \propto t^{-1/3}\) if we
take \( n=1\); thus, we can estimate the initial bulk Lorentz factor
\(\Gamma_{0}\).  The result is \(\Gamma_{0} \sim (1-2) \times
(10^{6.5}/10^{0.5})^{1/3} = 100-200\), where we set the time of the
burst as \( 10^{0.5}\) s.  Kulkarni et al. (1998) also gave an
estimate of the mass of the relativistic shock \(M_{\rm ej} \sim
10^{-5} M_{\odot}\).  Then the initial energy of the blast wave is
estimated to be less than \( \Gamma_{0} M_{\rm ej} c^{2} \sim
(1.8-3.6) \times 10^{51}\) erg.

\section{discussion}

Neither of the X-ray transient sources detected by Narrow-Field
Instruments (NFI) on {\it BeppoSAX} coincides with SN 1998bw in their
positions.  One of them (1SAX J1935.0-5248) was reported to have a
constant flux, but the other one (1SAX J1935.3-5252) has been
fading(Pian et al. 1998a, 1998b).  If the latter was indeed an
afterglow of GRB 980425 and the burst occurred at the position, the
fireball model predicts an optical afterglow and a subsequent radio
afterglow with a detectable level of fluxes. Since the positions of
the two X-ray sources were 3' away from that of SN 1998bw, they would
have been visible there. The lack of such optical and radio afterglows
may also indicate that the sources had nothing to do with GRB 980425.

Kulkarni et al.(1998) reported the radio light curves in four
different wavelength bands(Figure 2).  If we closely look at the light
curves, we see that they show complicated structures before and around
their maximum epoch, but rather clean spectrum and temporal evolutions
later. The irregularities seen in the early light curves may possibly
be the result of inhomogeneity in external matter, absorption by dust
in the host galaxy, and radiative processes other than synchrotron
radiation.  The light curves are well fitted by power-law decays with
the same exponent, \(\delta' = - 1.67 \).  On the other hand, the flux
ratios between different wavelength bands give a spectral index
\(\beta = - 0.7\), which is close to the values reported by Kulkarni
et al.(1998). These requires \(n = 3.2\), which does not agree with
the value we chose to fit the light curves in Figure 1. Such a large
value of \(n\) leads to a negative \(t\), which may indeed suggest a
density inhomogeneity in external matter, although it is necessary to
do more detailed modeling of the radio light curves with a realistic
treatment of hydrodynamics including other potentially important
radiative processes(Nakamura et al. 1998).

\section{Conclusions}

We examined the optical and radio light curves of SN 1998bw and
\(\gamma\)- and X-ray fluxes observed in GRB980425.  Under the
hypothesis that the two events are of the same origin, we compare the
light curves that a simple fireball model predicts with these
observations.  As a result, we find the following interesting facts
that favor the possibility that GRB 980425 was physically related to
SN 1998bw.

\noindent
1.  The radio emission from SN 1998bw and the \(\gamma\)-rays (BATSE,
{\it BeppoSAX}) from GRB980425 can be interpreted as a single event
based on a simple fireball model, although the X-ray flux reported by
{\it BeppoSAX} team is a bit too large and hard to be reconciled with
a model as simple as the one adopted here.

\noindent
2. The X-ray flux predicted by the model is only marginally detectable
at the epochs {\it BeppoSAX} NFI was pointed to the field, which is
consistent with the nondetection of any X-ray afterglow in
GRB980425. The predicted X-ray flux falls below the detection limit of
{\it ASCA} in \(\sim 2-3\) months; therefore, it is not likely that we
can detect decaying X-ray emissions from the burst by further
observations.

\noindent
3. Observations suggest that there seems to be a {\it relativistic
blast wave} and a subrelativistic {\it bulk mass ejection} in SN
1998bw.  The former might have caused the \(\gamma\)-ray emission from
GRB980425 and the radio emission from SN 1998bw, while the latter
corresponds to the optical emission from SN 1998bw. This picture is
consistent with the hypernova or collapsar scenarios for \(\gamma\)-ray
bursters, which involve the collapse of a massive star.

Although the arguments here are not strong enough to claim the
physical association between SN 1998bw and GRB 980425, it is worth
noticing that the relativistic shock required to explain the radio
emission from SN 1998bw could be interpreted as a decelerated
``fireball'' that initially had a higher bulk Lorentz factor and was
able to radiate \(\gamma\)-rays observed in GRB980425.

\bigskip

The author would like to thank Drs. Ken'ichi Nomoto, Elena Pian, Stan
Woosley, Takayoshi Nakamura, Hideyuki Umeda, Toshio Murakami, and
Timothy Young for useful discussion.  He is also grateful to the
anonymous referee for useful comments to improve the manuscript and
the figures.  This work has been supported in part by the grant-in-Aid
for Scientific Research (05242102, 06233101) and COE research
(07CE2002) of the Ministry of Education, Science, and Culture in
Japan, and the fellowship of the Japan Society for the Promotion of
Science for Junior Scientists(6728).

\vskip 1cm

\noindent
Note.1: The late-time optical light curve of model CO138 is calculated
by extending the exponential decay of the light curve of the same
model as presented in Iwamoto et al. (1998).  The previous version of
the light curve was not accurate enough, and has been replaced by a
correct one.

\vskip 1cm

\noindent
Note.2: The positions of the two NFI X-ray sources have been revised
(GCN Circ. 155, which says that ``one of the NFI sources was at 50''
from SN 1998bw and therefore consistent with it'').  A following GCN
Circular (No.158) reported that the flux from the 'consistent source'
has shown a moderate decay ( approximately by a factor of two from
April-May to November in 1998). The slow decline could be a possible
sign of circumstellar interaction in SN 1998bw(Note that the last
statement is merely our guess yet).

\begin{table}
\vskip 0.2cm
\label{tabdd}
\begin{center}
\begin{tabular}{cccc}
\hline Time & Energy Band (keV) & Flux (or Fluence) & Source \\ \hline 0-25
s & 24-1,820  & \((5.5 \pm 0.7) \times 10^{46}\) erg s$^{-1}$ &
CGRO BATSE$^{1}$ \\ 0-30 s & 2-28  & 3 Crab & BeppoSAX WFC No.2$^{2}$ \\
\(\sim\) 1 d & 2-10  & \(\sim (1.6 \pm 0.3) \times 10^{-13}\) erg
s$^{-1}$ cm$^{-2}$ & BeppoSAX NFI$^{2,3}$ \\ \(\sim 1.92\) d & 2-10  & \(< 1.2 \times
10^{-13}\) erg s$^{-1}$ cm$^{-2}$ & BeppoSAX NFI$^{2,3}$ \\ 
\(\sim\) 6 d & 2-10  & \( < 1.0 \times 10^{-13}\)  erg s$^{-1}$ cm$^{-2}$ &
BeppoSAX NFI$^{2,3}$ \\
\hline
\end{tabular}
\caption{$^{1}$quoted from Galama et al. (1998), $^{2}$Pian et al. (1998),
$^{3}$ 1SAX J1935.3-5252 }
\end{center}
\end{table}

\newpage

\begin{figure}[ht]
\centerline{\psfig{figure=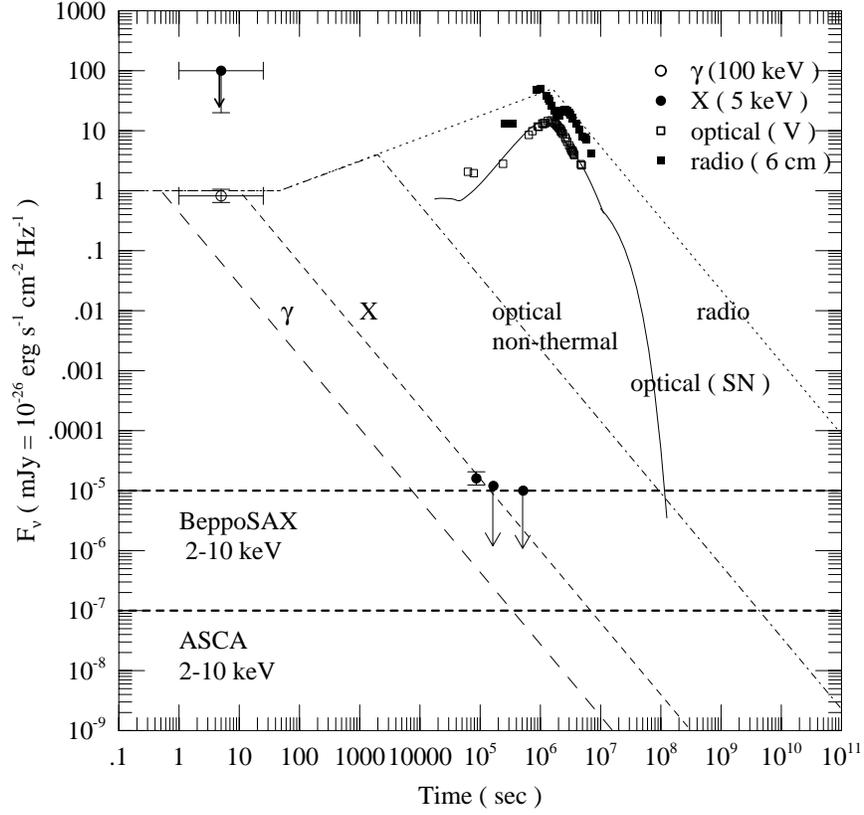,width=14cm}}
\caption[rv2.ps]{Light curves in \(\gamma\)-ray, X-ray, optical, and
radio wavelength bands predicted by a fireball model.  The detection
limits of {\it BeppoSAX} and {\it ASCA} satellites are indicated with
bold dashed lines. The optical light curve of SN 1998bw and its model
CO138(Iwamoto et al. 1998, see Note.1) are also plotted.  }
\end{figure}

\begin{figure}[ht]
\centerline{\psfig{figure=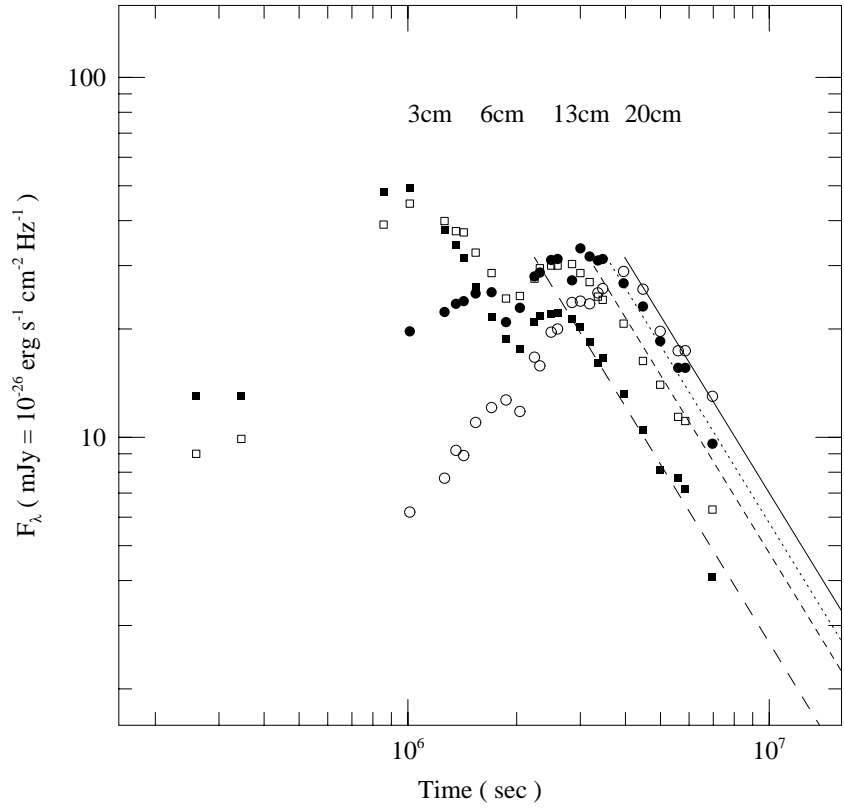,width=14cm}}
\caption[rv2.ps]{Radio light curves of SN 1998bw and at
wavelengths, 3cm(filled squares), 6cm(open squares), 13cm(filled
circles), and 20 cm(open circles) (Kulkarni et al. 1998).  The early
parts of the light curves show irregularities in spectral and temporal
evolutions. However, the later decaying parts are well reproduced by
power law decay curves with \(\delta' = -1.67, \beta = - 0.7\).
Solid, dotted, short-dashed, and long-dashed lines correspond to 20,
13, 6, and 3 cm bands.  }
\end{figure}


\begin{thebibliography}{}


\bibitem[Blandford \& McKee 1976]{Blandford-1976}
Blandford, R.D., \& McKee, C.F. 1976, Phys. of Fluids, 19, 1130

\bibitem[Blandford \& Znajek 1977]{Blandford-1977}
Blandford, R.D., \& Znajek, R.L. 1977, MNRAS, 179, 433

\bibitem[Galama et al. 1998]{Galama-1998}
Galama, T.J., et al. 1998, Nature, 395, 670

\bibitem[Iwamoto et al. 1998]{Iwamoto-1998}
Iwamoto, K., et al. 1998, Nature, 395, 672

\bibitem[Kulkarni et al. 1998]{Kulkarni-1998}
Kulkarni, S.R., et al. 1998, Nature, 395, 663

\bibitem[MacFadyen \& Woosley. 1998]{Woosley-1998} 
MacFadyen, A., \& Woosley, S.E. 1998, ApJ submitted, University of
California Observatories Preprint No.121

\bibitem[M\'esz\'aros \& Rees 1997]{Meszaros-Rees-1997}
M\'esz\'aros, P., \& Rees, M.J. 1997, ApJ, 476, 232

\bibitem[Nakamura et al. 1998]{Nakamura-1998} Nakamura, T., Iwamoto,
K., \& Nomoto, K. 1998, in preparation

\bibitem[Paczynski 1998]{Paczynski-1998}
Paczy\'nski, B. 1998, ApJ, 494, L45

\bibitem[Pian 1998]{Pian-1998} Pian, E., et al. 1998a, GCN Circ. 61 \\
(http://gcn.gsfc.nasa.gov/gcn/gcn3/061.gcn3)

\bibitem[Pian 1998]{Pian-1998}
Pian, E., et al. 1998b, GCN Circ. 69
\\
(http://gcn.gsfc.nasa.gov/gcn/gcn3/069.gcn3)

\bibitem[Piran 1997]{Piran-1997} Piran, T. 1997, in Some Unsolved
Problems in Astrophysics (eds. Bahcall, J. N. \& Ostriker, J.P.), 343

\bibitem[Rees \& M\'esz\'aros 1998]{Rees-Meszaros-1998}
Rees, M.J., \& M\'esz\'aros, P. 1998, ApJ, 496, L1

\bibitem[Wijers, Rees, \& M\'esz\'aros 1997]{Wijers-Rees-Meszaros-1998}
Wijers, R.A.M.J., Rees, M.J., \& M\'esz\'aros, P. 1997, MNRAS, 288, L51


\bibitem[Woosley et al. 1998]{Woosley-1998}
Woosley, S.E., Eastman, R.G., \& Schmidt, B.P. 1998, ApJ in press
(astro-ph/9806299)

\end{thebibliography}
\end{document}